\documentclass[12pt]{article}
\usepackage{amsfonts,amsbsy}

\newcommand{\be}{\begin{equation}}
\newcommand{\ee}{\end{equation}}

\newcommand{\bea}{\begin{eqnarray}}
\newcommand{\eea}{\end{eqnarray}}
\input{epsf.tex}

\begin{document}

\vskip -4cm

\begin{flushright}
FTUAM-98-20

IFT-UAM/CSIC-98-25
\end{flushright}

\vskip 0.2cm

{\Large
\centerline{{\bf On Nahm's transformation  }}
\centerline{{\bf   with twisted boundary conditions}}
\vskip 0.3cm

\centerline{\qquad  
  A. Gonz\'alez-Arroyo  }}
\vskip 0.3cm

\centerline{Departamento de F\'{\i}sica Te\'orica C-XI}
\centerline{ and Instituto de F\'{\i}sica Te\'orica C-XVI,}
\centerline{Universidad Aut\'onoma de Madrid,}
\centerline{Cantoblanco, Madrid 28049, SPAIN.}
\vskip 10pt

\vskip 0.8cm

\begin{center}
{\bf ABSTRACT}
\end{center}
Following two different tracks, we arrive at a definition of
Nahm's transformation valid for self-dual fields on the
4-dimensional torus $T^4$ with non-zero twist tensor.
The  transform is again a self-dual gauge field defined on a
new torus $\widehat{T}^4$ and with non-zero twist tensor. It preserves 
the property of being an involution.
\vskip 1.5 cm
\begin{flushleft}

\end{flushleft}

\newpage

\section{Introduction}
Self-dual SU(N) gauge fields are important objects both for
Mathematics and Physics. Polyakov\cite{polyakov} explained
their relevance in Quantum Field Theory shortly before the
instanton solution appeared\cite{BPST}. Not too long after,
the problem of constructing all self-dual configurations compactified on
a sphere was reduced to an algebraic problem in quaternions\cite{ADHM}
 (the ADHM construction).
On physical grounds, the compactification on a sphere amounts to the
condition of finite total action. This  restriction is,  from the physical
point of view, unrealistic.  Since the action is an extensive quantity, 
it is more sensible to demand finite
action density. For example, in computing the instanton contribution
to the path integral one deals with a finite density of instantons. The
isolated nature of these solutions, allows one to construct an approximate
finite density solution in terms of individual  instantons: the dilute gas
approximation. However, there is no rigorous justification for diluteness,
except for narrow instantons ($\rho \ll \Lambda_{\mbox{QCD}}$). It is, therefore,
interesting to  consider self-dual configurations with  finite action
density directly. From the mathematical point of view one could consider 
self-dual Yang-Mills fields on the torus. These configurations can be looked 
at as configurations in $R^4$ which are periodic in space-time. Of course the 
condition of periodicity is also unphysical, but could provide a starting 
point around which to construct a more realistic image of the Yang-Mills
vacuum. Furthermore, it is a simplification which might allow an analytic 
solution to be found. Putting Yang-Mills fields on the torus provides an 
additional richness in the form of new topological sectors. These are the 
twist sectors introduced by `t Hooft\cite{thooft}, which have an appealing 
physical interpretation in terms of electric and magnetic flux sectors. 
Unfortunately, the construction of self-dual gauge fields on the torus has 
proved a much harder task. Only a peculiar zero-measure set of these
configurations has been analytically constructed\cite{stsf,vb1}.

One important tool to study SU(N) self-dual configurations on the torus is
the Nahm transform\cite{nahm}. This is a mapping from self-dual
configurations to self-dual configurations, which is an involution.
It has the magic property of exchanging the rank of the group with the
topological charge. It is remarkable that one can see the ADHM construction
as a particular case of the Nahm transform. It would not be surprising if
the Nahm transform would hold the key to the construction of self-dual gauge
fields on the torus. Actually, it has been successfully used to construct
self-dual solutions on $R^3 \times S_1$\cite{kraanvanbaal,lee}.

However, the construction of the Nahm transform is based on the solution of the Weyl  equation
for fields in the fundamental representation of $SU(N)$. These fields are
only well defined if the twist sector is trivial. Henceforth, only for this
case  can one apply the standard construction. It is the purpose of
this paper
to show  how to construct the Nahm transform of self-dual gauge fields
which are periodic under a gauge transformation  belonging  to a
non-trivial twist sector. Traditionally several methods have been used to
deal with spinor fields in the fundamental representation in the presence
of twist. The first one is to go to $U(N)$ by adding a U(1) gauge field which 
will compensate for the non-trivial twist sector. We will leave this method 
out of our analysis and will concentrate on the other two approaches. The
second one is to add flavour\cite{cohen-gomez}, which  allows to define
boundary conditions which are consistent for the spinor fields. 
The last one, is to enlarge the torus size by considering  more than one 
cell of the original torus, in a way such that on the  new torus the twist 
sector is trivial(see for example Ref.~\cite{ntl}). In the next two sections
we will study the implementation  of these two methods.  Then in Section 4
we will relate the Nahm transforms obtained from the same gauge field, but 
considered as defined in different tori. Finally, the paper is closed with 
 a few concluding remarks.

\section{The Nahm transform for non-zero twist}

Let us start with $\mathbf{R^4}$ with Euclidean metric. Now, we introduce a
rank four lattice of vectors $\Lambda$. The generators are the vectors
$e^{(\mu)}$ whose components define an invertible matrix $M_{\mu \nu}=
e^{(\mu)}_{\nu}$. The quotient $T^4=\mathbf{R^4}/\Lambda$ defines a 4-dimensional
torus. Now let us consider an $SU(N)$ self-dual gauge field $A_{\mu}(x)$
living in $T^4$. This is equivalent to a gauge field on $\mathbf{R^4}$ 
and satisfying the following periodicity property:
\be
\label{tbc}
A_{\nu}(x+e^{(\mu)})= [\Omega_{\mu}(x)]A_{\nu}(x) \equiv  
 \Omega_{\mu}(x)\, A_{\nu}(x)\, \Omega^{\dagger}_{\mu}(x) +
 \imath\ \Omega_{\mu}(x)\,  \partial_{\nu} \Omega^{\dagger}_{\mu}(x)\ \ ,
\ee
where the matrices $\Omega_{\mu}(x)$ belong to $SU(N)$. They must satisfy the
following consistency condition:
\be
\label{twist}
\Omega_{\mu}(x+e^{(\nu)})\, \Omega_{\nu}(x)= z_{\mu \nu}\,
\Omega_{\nu}(x+e^{(\mu)})\, \Omega_{\mu}(x)\ \ .
\ee
The constants  $z_{\mu \nu}$ are elements of  $Z(N)$ (the $SU(N)$ center),
and, hence,  can 
be written as follows:
\be
z_{\mu \nu} = \exp(2 \pi \imath \frac{n_{\mu \nu}}{N})\ ,
\ee
where $n_{\mu \nu}$ is an antisymmetric tensor of integers $\bmod\,  N$ ({\em the 
twist tensor}). 
With an appropriate choice of generators the twist tensor takes the form
\be
\label{canonical}
\mathbf{n}= \left(  \begin{array}{cc}
0 &
{\bf \Xi} \\ - {\bf \Xi} & 0  \end{array}   \right)\ ,
\ee
where $\mathbf{\Xi}= \mbox{diag}(q_1,q_2)$ and $q_1$,  $q_2$ are positive
integers ($1 \leq q_i \leq N$).

We may now introduce the following family of $U(N)$ self-dual gauge fields
on the torus:
\be
A^z_{\mu}(x)= A_{\mu}(x) + 2 \pi  z_{\mu}\, \mathbf{I}\ .
\ee
The arbitrary real numbers $z_{\mu}$ can be  arranged into a four-vector $z$, 
and $\mathbf{I}$ is the $N \times N$
identity matrix. The field strength,  and the topological charge $Q$ are
independent of $z$. 

 Let us now consider an orthonormal basis of the space of  solutions of 
 the Weyl equation in the fundamental
 representation with $A^z_{\mu}(x)$ as background field:
 \be
 \label{weyl}
 \overline{D}^z\Psi^{\alpha}(x,z)=0\ \ .
 \ee
 The positive chirality Weyl operator  is given by $\overline{D}^z \equiv D^z_{\mu} \bar{\sigma}_{\mu}$
  ($D^z_{\mu} = \partial_{\mu} - \imath A^z_{\mu}(x)$),
where   $\bar{\sigma}_{\mu}=(\mathbf{I_{2\times 2}}, \imath\, \vec{\sigma})$
and $\sigma_i$ are the Pauli matrices.

Now let $e(n)$ denote an element of the lattice $\Lambda$:
\be
e(n)=n_{\mu}e^{(\mu)}\ , 
\ee
where $n$ is a vector of integers. To this element we may associate an $SU(N)$
matrix $\Omega_n(x)$. These
  matrices   can be obtained by
composing in a given order the matrices $\Omega_{\mu}(x)$ $n_{\mu}$ times.
Changing the order of multiplications could change the  $\Omega_n(x)$ by 
multiplication by an 
element of the center. Once a choice is made,  we can write:
\be
A_{\nu}(x+e(n))= [\Omega_n(x)]A_{\nu}(x) \ \ .
\ee

Let us first assume that the twist is trivial $n_{\mu \nu} = 0$.
Then, by virtue of the Atiyah-Singer index theorem~\cite{AS} and the
assumption  that there are  no negative chirality zero-modes, one can find 
$Q$ orthonormal solutions of the Weyl equation~(\ref{weyl})
$\Psi^{\alpha}(x,z)$ ($\alpha=1,\ldots, Q$) satisfying:
\be
\label{bc}
\Psi^{\alpha}(x+e(n),z)= \Omega_n(x)\, \Psi^{\alpha}(x,z)
\ee
The previous equation is consistent because the  $\Omega_n(x)$ commute
among themselves.  From these solutions, we can construct the Nahm transform
of $A_{\nu}(x)$ as follows:
\be
\label{nahmeq} 
(\widehat{A}_{\mu}(z))_{\alpha \beta}=\imath \int_{T^4} d^4x\; \Psi^{\dagger \alpha
}(x,z)\,
\frac{\partial}{\partial z_{\mu}}  \Psi^{\beta}(x,z) 
\ee
It can be shown, that $\widehat{A}_{\mu}(z)$ is an $SU(Q)$ 
self-dual gauge field
with topological charge $N$~(See Ref.~\cite{PB2} for a proof of these and 
other properties of the Nahm transform).
Now, let us investigate the periodicity properties of $\widehat{A}_{\mu}(z)$.
For that we notice  that $\exp(- 2 \pi \imath \tilde{z}_{\mu} x_{\mu})\,
\Psi^{\alpha}(x,z+\tilde{z})$ satisfies the same equation than
$\Psi^{\alpha}(x,z)$. The boundary conditions are, however, different 
in general. In order to satisfy Eq.~\ref{bc}, one must have that $\tilde{z}_{\mu}
e^{(\nu)}_{\mu} \in \mathbf{Z}$. This is the condition that defines the dual
lattice $\widetilde{\Lambda}$. The  basis dual to  $\{e^{(\nu)}\}$ is given by
$\{\tilde{e}^{(\nu)}\}$ satisfying:
\be
\tilde{e}^{(\mu)}_{\rho}\, e^{(\nu)}_{\rho}\, = \delta_{\mu \nu}\ \ .
\ee
Now, since $\Psi^{\alpha}(x,z)$ is a basis of the space of solutions 
with boundary conditions~(\ref{bc}), we must have:
\be
\Psi^{\alpha}(x,z+\tilde{e}^{(\mu)})=  \Psi^{\beta}(x,z)
(\widehat{\Omega}^{\dagger}_{\mu}(z))_{\beta \alpha}\, \exp(2 \pi \imath\,
\tilde{e}^{(\mu)}_{\rho} x_{\rho}) \ .
\ee
Then one sees that $\widehat{A}_{\nu}(z)$ satisfy boundary conditions 
similar to
those of equation~(\ref{tbc}) with $\Omega_{\mu}(x)$ replaced by
$\widehat{\Omega}_{\mu}(z)$. In summary, the Nahm transformed field lives in the
dual torus $\widetilde{T}^4=\mathbf{R}^4/\widetilde{\Lambda}$.

Let us go back to the general twist case.  The previous procedure fails
because Eq.~(\ref{bc}) becomes inconsistent for non-trivial twist. The strategy
that we will put forward is the following. Consider a sublattice $\Lambda'$
of  $\Lambda$. The sublattice is chosen in such a way that for all $e(n)\in
\Lambda'$ the corresponding $\Omega_n(x)$ commute (Actually the $\Omega_n(x)$
must define a representation of $\Lambda'$). We will impose
Eq.~(\ref{bc}) only in $\Lambda'$. This is equivalent to
considering the fields as living in a wider torus  $T'^4=\mathbf{R}^4/\Lambda'$.
In this case the Nahm transform can be constructed in the same way as
described before, and lives in $\widetilde{T}'^4=\mathbf{R}^4/\widetilde{\Lambda'}$.
The rank of the Nahm transformed field is now $\widehat{N}=Q\,
|\Lambda/\Lambda'|$, where $|\Lambda/\Lambda'|$ is the order of the quotient
group of the two lattices. Henceforth, the basis of the space of
solutions is $\Psi^{\tilde{\alpha}}(x,z)$ with $\tilde{\alpha} =1, \ldots
,\widehat{N}$.

We want now  to explore the consequences  of the original stronger 
periodicity condition for the
fields. For any $n \in \mathbf{Z}^4$ we can define:
\be
\label{def1}
\Psi^{(n) \tilde{\alpha}}(x,z) =  \Omega^{\dagger}_n(x)\,
\Psi^{\tilde{\alpha}}(x+e(n),z)\ \ .
\ee
If $e(n)\in\Lambda'$, the boundary  conditions imply that the
$\Psi^{(n)\tilde{\alpha}}$
coincide with  $\Psi^{\tilde{\alpha}}$. Notice, however, that for any $n$
the previous functions are solutions of the Weyl equation~(\ref{weyl}).
The problem is that, if the twist tensor is non-zero,  the previous functions
satisfy boundary conditions which differ by multiplication by an element of
the center from those of $\Psi^{\tilde{\alpha}}(x,z)$. However, remember that
the same situation took place whenever we  displaced z by a vector
not belonging to the dual lattice and multiplied it by a corresponding phase.  
Henceforth, combining both operations, we arrive at our most 
important result, summarized in the
following formulas:
\begin{eqnarray}
\label{main}
&\Psi^{(n) \tilde{\alpha}}(x,z+\Delta) \exp(- 2 \pi \imath\, \Delta_{\mu}
x_{\mu}) =  \Psi^{\tilde{\beta}}(x,z)\,
(\widehat{\Omega}^{\dagger}_{n \Delta}(z))_{\tilde{\beta} \tilde{\alpha}}& \ \\
\label{cond}
&\frac{1}{N}n_{\mu \nu} n_{\nu} m_{\mu} - \Delta_{\mu} e(m)_{\mu}\, \in
\mathbf{Z}
\ \ \ \mbox{for}\  e(m) \in \Lambda' &           .
\end{eqnarray}
The second equation (\ref{cond}) is the necessary and sufficient condition 
for the left
hand side of Eq. (\ref{main}) to satisfy the same boundary conditions as
 $\Psi^{\tilde{\alpha}}(x,n)$. The first equation then expresses the fact
 that it can
be written as a linear combination of the basis of the space of solutions.
One can see from~(\ref{def1})  that the functions $\Psi^{(n)
\tilde{\alpha}}(x,z)$ are orthonormal irrespective of $n$, and hence the
$\widehat{\Omega}(z)$ are unitary.
Then,  one can conclude that for any $\Delta$ such that there exists an $n$
satisfying Eq.~(\ref{cond}) we have:
\be
\widehat{A}_{\nu}(z+\Delta)= [\widehat{\Omega}_{n \Delta}(z)]\, \widehat{A}_{\nu}(z)\ .
\ee
This formula is similar to Eq.~(\ref{tbc}), and is the main formula we were
looking for. For $\Delta$ belonging to $\widetilde{\Lambda'}$, it amounts to what
was known previously. But this formula makes the periodicity valid for a
larger lattice $\widehat{\Lambda}$. We must now study the consistency conditions
that follow from equation~(\ref{main}). If we consider two sets $(n,\Delta)$
and $(n',\Delta')$ satisfying Eq.~(\ref{cond}), we can deduce:
\begin{eqnarray}
\label{tbcc}
&\widehat{\Omega}_{n \Delta}(z+\Delta')\, \widehat{\Omega}_{n' \Delta'}(z)= \\
\nonumber
&\exp\{ \frac{2 \pi
\imath}{N} n_{\mu \nu} n_{\mu}  n'_{\nu} + 2 \pi \imath
(\Delta_{\rho}e(n')_{\rho} - \Delta'_{\rho}e(n)_{\rho})\}\ 
\widehat{\Omega}_{n' \Delta'}(z+\Delta)\, \widehat{\Omega}_{n \Delta}(z)\ .
\end{eqnarray}
This condition is the counterpart of Eq.~(\ref{twist}) for the Nahm transform.

Now, all we need is to straighten up  the conditions implied by Eqs.~
(\ref{main}-\ref{tbcc}). Let us stick to the form Eq.~(\ref{canonical}). Then,
if  $q_i \ne 0$ we can form the integers $p_i=N/\gcd(N,q_i)$. The case $q_i=0$
can be treated in the same fashion by setting $q_i=N$, $p_i=1$. The lattice
$\Lambda'$ is the one spanned by the vectors  $\{e^{(0)},e^{(1)},p_1 e^{(2)}, p_2 e^{(3)}\}$.
The elements of $\Lambda/\Lambda'$ can be labeled  by two integers
$n_2=0, \ldots, (p_1-1)$ and $n_3=0, \ldots, (p_2-1)$. Then an arbitrary
element\footnote{The corresponding element of $\Lambda/\Lambda'$ is given by
$n_2=s_1 k_0 \bmod p_1$ ; $n_3=s_2 k_1 \bmod p_2$, with $s_i$ defined after
Eq.~(\ref{nstar}).} of
the  lattice $\widehat{\Lambda}$
is given by $\Delta=\hat{e}^{ (\mu)} k_{\mu}$~($k_{\mu} \in
\mathbf{Z}$). 
The generators are given by:
\begin{eqnarray}
\nonumber
\hat{e}^{ (0)} = \tilde{e}^{ (0)}/p_1 & \\
\label{lambstar}
\hat{e}^{ (1)} = \tilde{e}^{ (1)}/p_2 &\\
\nonumber
\hat{e}^{ (2)} = \tilde{e}^{ (2)}/p_1 &\\
\nonumber
\hat{e}^{ (3)} = \tilde{e}^{ (3)}/p_2 &\ \ . 
\end{eqnarray}
From Eq.~(\ref{tbcc}) one can compute the twist matrix $\hat{n}_{\mu  \nu}$ 
associated to the 
Nahm transform:
\be
\label{nstar}
\mathbf{\hat{n}}=\left(  \begin{array}{cc}
0 &
{\bf \widehat{\Xi}} \\ - {\bf \widehat{\Xi}} & 0  \end{array}   \right)\ \ \mbox{with}\ {\bf
\widehat{\Xi}}= \mbox{diag}((p_1-s_1) p_2 Q\,,\,(p_2-s_2) p_1 Q)\ \ ,   
\ee
where $s_i$ is an  integer ($0 \leq s_i \leq (p_i-1)$) satisfying 
$s_i q_i =\gcd(q_i,N) \bmod N$. 
We have taken the rank of the Nahm
transform $\widehat{N}$ to be $Q\, p_1 p_2$. Taking into account that the topological
charge $Q$ has the form $l-\frac{q_1 q_2}{N}$ with $l$ an integer (see for
example Ref.~\cite{torusrev}) one can easily show  that the 
$\hat{n}_{\mu  \nu}$ are 
integers. 

To conclude this section, we mention that if we apply  Nahm's transformation
to the Nahm transform, we obtain back the original gauge field with twist
in $T^4$. If we label with a $\wedge$  all quantities referring to the 
Nahm transform, one can derive the following relations:
\begin{eqnarray}
\nonumber  \widehat{N}= Q p_1 p_2 \ \ & \ \ \widehat{Q}= \frac{N}{p_1 p_2} \\
 \hat{p}_i= p_i \ \ & \ \ \hat{s}_i=p_i- \frac{q_i}{\gcd(N,q_i)}\ .
\end{eqnarray}
With these relations  it is not hard to show that
$\widehat{\widehat{\Lambda}}=\Lambda$,
and that after applying the Nahm transform twice we recover the original
set of parameters. It is also easy to show that the Nahm transform of a
gauge field configuration with non-orthogonal twist ( $\frac{1}{8}
\epsilon^{\mu \nu \rho \sigma}n_{\mu \nu} n_{\rho \sigma} \neq 0 \bmod N$)
has also non-orthogonal twist.

\section{Adding flavour}
Now let us pursue a different  strategy. What we will do is to map our
$SU(N)$ gauge field into an $SU(N \, N_0)$ one as follows:
\be
A_{\mu}(x) \longrightarrow A_{\mu}(x)\otimes \mathbf{I}_{N_0 \times N_0}\ \ .
\ee
The advantage of this construction is that one can choose twist matrices
in $SU(N \, N_0)$ which belong to the trivial  twist sector:
\be
\Omega_{\mu}(x) \longrightarrow \Omega_{\mu}(x) \otimes \Gamma_{\mu}\ \ ,
\ee
where $\Gamma_{\mu}$ are constant $SU(N_0)$ twist matrices (in some cases
 we will
need to relax them to being in $U(N_0)$) satisfying:
\be
\label{twisteat}
\Gamma_{\nu} \Gamma_{\mu} =  z_{\mu \nu}\, \Gamma_{\mu} \Gamma_{\nu}\ \ .
\ee
 The previous condition is required to cancel the twist of the $\Omega_{\mu}$.
The solutions are the so-called {\em twist eaters}. These exist provided
$N_0$ is an integer multiple of $p_1 p_2$ (see Ref.~\cite{torusrev}-\cite{twisteat} and
references therein for the existence and properties of twist eaters).
Let us take precisely $N_0=p_1
p_2$, where the solution is irreducible. Then we have produced an $SU(p_1
p_2 N)$ self-dual gauge field having topological charge $Q p_1 p_2$ and
transforming under translations by elements of $\Lambda$ with twist matrices
belonging to the trivial twist sector. Hence, we are in the conditions under
which we can apply the standard Nahm transform procedure, thus
generating an $SU(Q  p_1 p_2)$ self-dual gauge field having topological charge
$N p_1 p_2$ and living in $\widetilde{T}^4$. This Nahm transform coincides with the 
one constructed in the last section with a different procedure. 

Before proving this, we will show  that the new Nahm transform field is periodic
(modulo gauge transformations) not only in the torus $\widetilde{T}^4$, but in
the finer one $\widehat{T}^{ 4}$. The proof is as follows. Let us consider a basis of
the space of solutions of the Weyl equation with the appropriate boundary
conditions $ \Phi^{i \alpha}(x,z)$. The index $i$ is the flavour index
($i=1,\ldots N_0$), and the index $\alpha$ runs over the $Q p_1 p_2$ linearly
independent solutions. For every $i$ and $\alpha$ the previous fields satisfy 
the Weyl equation~(\ref{weyl}). Furthermore, they satisfy the following 
boundary condition:
\be
\label{bc2}
\Phi^{i \alpha}(x+e(n),z)= \Omega_n(x)\, \Gamma_n^{i j}\, \Phi^{j
\alpha}(x,z)\ . 
\ee
 The constant unitary $N_0 \times N_0$ matrices $\Gamma_{n}$ are constructed in the same way as
the $\Omega_n(x)$. The product  $\Omega_n(x)\, \Gamma_n$ does not depend  on
the product sequence  taken to construct them. Now, let us show the 
required periodicity
in $\widehat{T}^{ 4}$, by constructing the following functions:
\be
\widetilde{\Phi}^{i \alpha}(x,z+\Delta)= \exp(2 \pi \imath\, \Delta_{\mu} x_{\mu})\,
S_{\Delta}^{i j }\,   \Phi^{j \alpha}(x,z)\ \ ,
\ee
where the $S_{\Delta}$ are constant unitary matrices. The functions $\tilde{\Phi}$
satisfy the Weyl equation for $z + \Delta$. Furthermore, they satisfy the 
boundary conditions (\ref{bc2}) provided:
\be
\label{condition}
S_{\Delta}\, \Gamma_n \, S_{\Delta}^{\dagger} = \exp(- 2 \pi \imath\, \Delta_{\mu}
e(n)_{\mu})\; \Gamma_n  \ \ .
\ee
The set of values of $\Delta$ for which there exist a constant unitary 
$S_{\Delta}$ satisfying the previous equation defines a new lattice, 
which obviously  contains $\widetilde{\Lambda}$. We have now to show that it 
is precisely $\widehat{\Lambda}$. 

Let us write:
\be
S_{\Delta}=\Gamma_0^{l_0}\, \Gamma_1^{l_1}\, \Gamma_2^{l_2}\,
\Gamma_3^{l_3}\  \ ,
\ee
with $l_i$  some integers depending on $\Delta$. The
condition~(\ref{condition}) now reads:
\be
\label{def2}
\frac{n_{\mu \nu} n_{\mu} l_{\nu}}{N} +
\Delta_{\mu} e(n)_{\mu}\, \in \mathbf{Z}\ \ .  
\ee
This condition must hold for all values of the integers $n_{\mu}$. It is
easy to see that this has a solution if and only if $\Delta \in
\widehat{\Lambda}$.
Actually Eq.~(\ref{def2}) represents a much cleaner characterization
of $\widehat{\Lambda}$ than Eq.~(\ref{cond}).
The Nahm transform defined in this way 
coincides  both in the rank and in the defining torus with the one
constructed in the previous section. In the following paragraph we will 
prove that  the two procedures do indeed give the same Nahm transformed
field.

For that purpose, we take the explicit form
of the twist tensor~(\ref{canonical}) and  make a particular choice of the $\Gamma_{\mu}$
matrices. $\Gamma_0$ and  $\Gamma_1$ can be chosen diagonal, while the
other two are necessarily non diagonal. However, $\Gamma_2^{p_1}$ and
$\Gamma_3^{p_2}$ are also diagonal. It is also possible, if we take the
matrices to be in $U(N_0)$ rather than $SU(N_0)$, to choose these
4 diagonal matrices such that the $1,1$ component is 1 (see
Ref.~\cite{torusrev}-\cite{twisteat} and references therein for a proof of these facts). 
The advantage of this
choice is that now the functions $\Phi^{1 \alpha}(x,z)$ satisfy the same
conditions as the functions $\Psi^{\tilde{\alpha}}(x,z)$ introduced in the
previous chapter. Clearly the former can be written as linear combinations
of the latter or vice versa. Notice that the difference between both
constructions is that in one of them($\Phi$) the functions have a flavour index
taking $N_0=p_1p_2$ values, in the other case ($\Psi$) the functions are
taken to live in a larger torus $T'^4$ rather than $T_4$. We will now see how
both points are related. Translating $\Phi^{1 \alpha}(x,z)$ we get:
\be
 \Phi^{1 \alpha}(x+n_2 e^{(2)}+n_3 e^{(3)},z)=\Omega_n(x)\, \Gamma_n^{1 j}\,
 \Phi^{j \alpha}(x,z)\ \ ,
\ee
where $n=(0,0,n_2 ,n_3)$. Now, it can be seen that $\Gamma_n^{1 j}$  is only
non-zero for one particular j. This establishes a one  to one correspondence
between flavour components and elements $e(n)\in \Lambda/\Lambda'$. Even
more the  $\Gamma_n^{1 j}$ can be chosen to be $1$ (with $\Gamma$ in $U(N_0)$). 
These relations serve
to fix normalizations, and with an appropriate choice of basis in both cases
one can obtain the following relation:
\be
\Phi^{j \alpha}(x,z) = \Omega_{-n}(x)\, \Psi^{\alpha}(x+n_2 e^{(2)}+n_3
e^{(3)},z) 
\ee
which establishes the connection between the formalism of the previous section
and the one of this section. One can easily see that the Nahm transformed
fields constructed with both methods coincide. In the method of last section
one has no flavour index, but the integral necessary to construct the
Nahm transform is in $T'^4$ which is a torus $p_1 p_2$ times larger than $T^4$.
In summary, replicas can be looked at as flavours.

\section{Replicas without twist }
We will now apply the formalism of section 2 to understand what happens
when we replicate the torus in the absence of twist. We start with the
self-dual $SU(N)$ gauge field $A_{\mu}(x)$ defined on the torus
$T^4=\mathbf{R^4}/\Lambda$. The topological charge is $Q$. Now we can
construct the Nahm transform, which is an $SU(Q)$ self-dual gauge field
defined in $\widetilde{T}^4=\mathbf{R^4}/\widetilde{\Lambda}$. 
On the other hand,
we might look at  $A_{\mu}(x)$ as a gauge field defined on the larger
torus $T'^4=\mathbf{R^4}/\Lambda'$, where $\Lambda'$ is a sublattice of
$\Lambda$. Applying the Nahm transform in this case gives an
$SU(Q\,|\Lambda/\Lambda'|)$ gauge field living in the dual torus
 $\widetilde{T}'^4=\mathbf{R^4}/\widetilde{\Lambda'}$. The question that we want to 
 answer is what is the relation between both Nahm transforms. 

 Without loss of generality we can restrict ourselves to the case in which we
 replicate the torus in one direction. By repeated application of this
situation we can deal with the general case. Let 
$e^{(\mu)}$ be the generators of $\Lambda$. The lattice $\Lambda'$ has the
same generators except for $e'^{(0)}=L\, e^{(0)}$ where $L=|\Lambda/\Lambda'|$ is a positive integer.
For the dual tori we have $\tilde{e}'^{(0)}=\frac{1}{L}\, \tilde{e}^{(0)}$.
Now consider the solutions of the Weyl equation in $T^4$. There are $Q$
orthonormal linearly independent solutions $\Psi^{\alpha}(x,z)$ satisfying the boundary
conditions Eq.~(\ref{bc}). With them we can construct the Nahm transform as
usual. Now let us consider the following functions:
\be
\Psi^{\alpha k}(x,z)=\Psi^{\alpha}(x,z+\frac{k}{L}\tilde{e}^{(0)})\, \exp(-
2 \pi \imath\, x_{\mu}\tilde{e}^{(0)}_{\mu} \frac{k}{L})\ \ ,
\ee
where $k$ is an integer ranging from $0$ to $L-1$. They are all solutions of
the Weyl equation. Let us see how these
functions behave under translations along $e^{(0)}$. We have 
\be
\Psi^{\alpha k}(x+l\,e^{(0)},z)= \exp(-2 \pi \imath \frac{k l}{L})\
\Omega_{n_l}(x)\,
\Psi^{\alpha k}(x,z)\ ,
\ee
where $n_l=(l,0,0,0)$.
Notice that, although only for $k=0$  the boundary conditions on $T^4$
are satisfied, they all satisfy the right boundary conditions on $T'^4$, i.e.
for $l=L$. Since there are $LQ$ different functions, they constitute a basis of
the space of solutions provided they are linearly independent. We will show
that indeed they are all mutually orthogonal in $T'^4$. The proof goes as
follows:
\begin{eqnarray}
\nonumber
&\int_{T'^4} dx\, \Psi^{\dagger \alpha' k'}(x,z)\,  \Psi^{ \alpha k}(x,z)=\\
\label{seq}
&\sum_{l=0}^{L-1} \int_{T^4} dx\, \Psi^{\dagger \alpha' k'}(x+l\,e^{(0)},z)\,  \Psi^{ \alpha
k}(x+l\,e^{(0)},z)=\\
\nonumber
&\left(\sum_{l=0}^{L-1} \exp(2 \pi \imath \frac{l(k-k')}{L})\right)
\int_{T^4} dx\, \Psi^{\dagger
\alpha' k'}(x,z)\, \Psi^{ \alpha k}(x,z)=\\
\nonumber
&L\, \delta_{k k'}\, \delta_{\alpha \alpha'}
\end{eqnarray}
In the last identity we used the orthogonality of the $\Psi^{\alpha}(x,z)$
for arbitrary $z$. 

Dividing the functions $\Psi^{ \alpha k}(x,z)$ by $\sqrt{L}$, we have the
necessary orthonormal basis of the space of solutions that we need to
construct the Nahm transform in $T'^4$. Applying the definition of the Nahm 
transform, and after similar manipulations as in Eq.~(\ref{seq}) we arrive at:
\be
\left(\widehat{A}(z)\right)_{(\alpha k)(\alpha' k')}= \delta_{k k'}\,
\left(\widehat{A}(z+\frac{k}{L}\tilde{e}^{(0)})\right)_{\alpha
\alpha'}\ \ ,
\ee
where the left hand side is the Nahm transform defined in $T'^4$ and it is
expressed in terms of the Nahm transform in $T^4$. It is clear from our
formulas that the former is reducible and invariant (modulo gauge
transformations) under translation of z by $\frac{1}{L}\tilde{e}^{(0)}$.

\section{Concluding remarks}
Let us first summarize our results. In the previous sections,
we have explained how to define the Nahm transform of an
$SU(N)$ self-dual gauge field living in the torus $T^4=\mathbf{R^4}/\Lambda$
with  twist tensor $\mathbf{n}$. This is again an $SU(Q)$ self-dual gauge
field defined on the torus $\widehat{T}^{ 4}=\mathbf{R^4}/\widehat{\Lambda}$
with twist tensor $\mathbf{\hat{n}}$. The explicit expression of
$\widehat{\Lambda}$
and $\mathbf{\hat{n}}$ are given in  Eqs.~(\ref{lambstar}),(\ref{def2}),(\ref{nstar})
for a basis of $\Lambda$
such that $\mathbf{n}$ has the canonical form~(\ref{canonical}).
If we want to give the relations in an arbitrary basis we have to
write $ \frac{{\bf n }}{N} = Q R^{-1} + \Sigma$, where $Q$, $\Sigma$ and $R$
are integer matrices, and the first two are antisymmetric ($R$ is positive
definite). In the canonical
basis the matrix $R$ is $\mbox{diag}(p_1,p_2,p_1,p_2)$. If we change
from the canonical basis to other by means of the integer unit determinant
matrix $U$, then $R \longrightarrow U^{-t} R U^{t}$, where $U^{t}$ is the
transpose of $U$ and $U^{-t}$ the inverse transpose. We also demand
the existence of another integer matrix $\widehat{Q}$
such that $\widehat{Q} Q= -\mathbf{I} + \Sigma''R$, where $\Sigma''$ is 
an arbitrary integer matrix. Given this notation we can summarize the
Nahm transformed expressions  as follows:
\bea
\nonumber
\widehat{M} = R^{-1} \widetilde{M}\\
\frac{\mathbf{\hat{n}}}{\widehat{N}}= \widehat{Q} R^{-t} +\Sigma'
\eea
where $(\widehat{M})_{\mu \nu}=\hat{e}^{(\mu)}_{\nu}$ is the matrix
giving the coordinates of the basis of $\widehat{\Lambda}$, and
$\Sigma'$ in an arbitrary integer matrix.

 The formalism reduces
 to the standard one in the absence of twist. One has simply to
 set $R=\mathbf{I}$, or  $q_i=N\,;\,p_i=1$  in the appropriate formulas.

 There are some applications of the present formalism that one can think of.
 It is possible, for example, 
 to extend the results of van Baal on the Nahm transform of 
 constant field strength self-dual solutions~\cite{bukow} to  
 the case of non-zero twist. There are other questions which 
 can be investigated with our formulas, such as the occurrence 
 of fixed points of the Nahm transform. In a previous paper~\cite{ntl}
 we applied  our numerical approach to the Nahm transform, to a 
 case which turned out to be a good candidate for a fixed point\footnote{
 the word fixed point has to be understood modulo an orthogonal 
 transformation in space-time}. 
 Indeed, the expressions presented here are in agreement 
 with the conclusions of that paper. Conversely, the numerical 
 results of that paper can serve as an explicit case where one can see 
 the present formalism at work.

\section*{Acknowledgements}
The author wants to thank Pierre van Baal, Margarita 
Garcia Perez and Carlos Pena for useful conversations 
on the subject. Special thanks go for Pierre van Baal 
for useful suggestions and a critical reading of the 
manuscript.

\end {document}